\documentclass[fleqn,usenatbib]{mnras}

\usepackage{newtxtext,newtxmath}

\usepackage[T1]{fontenc}
\usepackage{soul}

\DeclareRobustCommand{\VAN}[3]{#2}
\let\VANthebibliography\thebibliography
\def\thebibliography{\DeclareRobustCommand{\VAN}[3]{##3}\VANthebibliography}

\usepackage{graphicx}	
\usepackage{amsmath}	
\usepackage{url} 
\usepackage{hyperref}

\title[Anisotropic correlation functions for central galaxies]{Anisotropic correlation functions as tracers of central galaxy alignments in simulations}

\author[ Rodriguez, F., Merchán, M., Artale, M. C. \& Andrews, M.]{
Facundo Rodriguez,$^{1,2}$\thanks{E-mail: facundo.rodriguez@unc.edu.ar (FR)}
Manuel Merch\'an$^{1,2}$\thanks{E-mail: manuel.merchan@unc.edu.ar (MM)}, M. Celeste Artale$^{3,4,5,6}$ and Moira Andrews$^{5}$
\\
$^{1}$ CONICET. Instituto de Astronomía Teórica y Experimental (IATE). Laprida 854, Córdoba X5000BGR, Argentina.\\
$^{2}$ Universidad Nacional de Córdoba (UNC). Observatorio Astronómico de Córdoba (OAC). Laprida 854, Córdoba X5000BGR, Argentina.\\
$^{3}$ Physics and Astronomy Department Galileo Galilei, University of Padova, Vicolo dell’Osservatorio 3, I-35122, Padova, Italy\\
$^{4}$ INFN - Padova, Via Marzolo 8, I-35131 Padova, Italy\\
$^{5}$ Department of Physics and Astronomy, Purdue University, 525 Northwestern Avenue, West Lafayette, IN 47907, USA \\
$^6$ Departamento de Ciencias Fisicas, Universidad Andres Bello, Fernandez Concha 700, Las Condes, Santiago, Chile
}

\date{Accepted XXX. Received YYY; in original form ZZZ}

\pubyear{2022}

\begin{document}
\label{firstpage}
\pagerange{\pageref{firstpage}--\pageref{lastpage}}
\maketitle

\begin{abstract}
Motivated by observational results, we use IllustrisTNG hydrodynamical numerical simulations to study the alignment of the central galaxies in groups with the surrounding structures. This approach allows us to analyse galaxy and group properties not available in observations.
To perform this analysis, we use a modified version of the two-point cross-correlation function and a measure of the angle between the semi-major axes of the central galaxies and the larger structures. 
Overall, our results reproduce observational ones, as we find large-scale anisotropy, which is dominated by the red central galaxies. In addition, the latter is noticeably more aligned with their group than the blue ones.
In contrast to the observations, we find a strong dependence of the anisotropy on the central galaxy with mass, probably associated with the inability of observational methods to determine them. This result allows us to link the alignment to the process of halo assembly and the well-known dependence of halo anisotropy on mass.
When we include the dark matter distribution in our analysis, we conclude that the galaxy alignment found in simulations (and observations) can be explained by a combination of physical processes at different scales: the central galaxy aligns with the dark matter halo it inhabits, and this, in turn, aligns with the surrounding structures at large scales.

\end{abstract}




\section{Introduction}


The shapes and angular momentum of dark matter halos are driven by accretion and tidal processes with the surrounding matter \citep{Ciotti1994, Usami1997,Porciani2002a, Fleck2003}. In this context, galaxies form and evolve in the potential well of dark matter halos and, in consequence, tend to follow a preferential shape, orientation and distribution \citep{Pen2000, Catelan2001,Crittenden2001, Porciani2002b,Jing2002,Ciotti1994, Usami1997,Porciani2002a, Fleck2003}. These features imprint information on the way the environment influences their formation history.

Observational evidence shows that galaxy alignments vary according to their colours, luminosity and star formation history. In particular, early-type/red galaxies tend to be  more aligned with the large-scale dark matter distribution as a consequence of their build-up of matter, than late-type/blue galaxies \citep[see, e.g.,][]{Sales2004,yang2006alignment,Agustsson2010,Kirk2015}. On the other hand, several works show that satellite galaxies tend to be distributed along the galactic plane of the central galaxies \citep{Yang2005,Wang2008,Libeskind2015,Welker2018,Pawlowski2018}, and red satellite galaxies are more aligned when the central one is red than when it is blue \citep[][]{Kiessling2015,Kirk2015, Johnston2019}. The dependence on the galaxy colours is referred as \textit{anisotropic quenching} and \textit{angular conformity} in several works \citep[see][and references therein]{Wang2008,Stott2022}, and it can be related to the \textit{galactic conformity} \citep[][]{Bray2016,Otter2020,Maier2022}. As mentioned above, the preferential alignments and galaxy occupancy in haloes can be explained as a manifestation of the large-scale environmental conditions in combination with baryonic processes such as stellar and AGN feedback.  

In a recent work, \cite{Rodriguez2022} study the alignment of the central galaxies of the groups with the surroundings, using spectroscopic data from the Sloan Digital Sky Survey Data Release 16 \citep[SDSS DR16,][]{ahumada2020} and the galaxy group finder presented in \cite{rodriguez20}. Their results show that central galaxies are aligned with the satellite galaxies within the same halo (which agrees with \citealt{yang2006alignment}) and with nearby structures up to a distance of $\sim$ 10 Mpc. Moreover, the galaxy alignment shows a strong dependence on the colour, where red central galaxies are more aligned with the environment than the blue ones. Other properties such as the ellipticity of the central galaxy and the  halo mass do not show a dependence with galaxy alignments. \cite{rodriguez20} also investigate the correlation function taking the group shape as a reference. They find, as expected, that the one-halo term show a large anisotropy, higher than that obtained using the shape of the central galaxy as a reference. 
On the other hand, the two-halo term is very similar in both amplitude and anisotropy as it is calculated with respect to the shape of the halo or the central galaxy.
However, the probability distribution function of the angle between the major axes of the central galaxy and that of its group shows an overall faint alignment dominated by the central red galaxies.

Based on the findings presented in \cite{Rodriguez2022}, in this work we use the cosmological hydrodynamical simulation IllustrisTNG\footnote{\url{http://www.tng-project.org}} (hereafter, simply TNG) to extend the analysis. 
This approach mainly allows for a three-dimensional analysis without projection effects. In addition, it enables us to take advantage of the dark matter distribution provided by the TNG suite to study other secondary dependencies of the dark matter halo besides the halo mass.
Cosmological simulations and in particular, hydrodynamical cosmological simulations, have proven to be a powerful tool to study the intrinsic alignments and spatial distribution of galaxies and dark matter haloes \citep[][]{Deason2011,Dong2014,Shao2016,Welker2018,Zjupa2020,Shi2021,Shi2021,Shi2021b,RagoneFigueroa2020,Samuroff2020}.

Here we use the cosmological hydrodynamical simulation IllustrisTNG. Previous works based on this simulation analyzed the alignments of satellite galaxies with respect to the central ones \citep{Tenneti2021}, and \citet{Jagvaral2022} report an alignment of the elliptical galaxies in two-halo term scales of the order of $1-4$~Mpc. Furthermore, \citet{Zhang2022} finds that galaxies with mass above $M_{*}>10^{10.5}$~M$_{\odot}$ present an excess of alignment between the galaxy shapes and the cosmic web filaments. Our results complement previous findings with the TNG simulation and we intend to investigate numerically the findings showed by \citet{Rodriguez2022}.  

This paper is organised as follows. Section~\ref{sec:TNG} provides a description of the TNG simulation box and the halo and galaxy properties analysed. Section~\ref{cf} presents the results for the anisotropic correlation function for central galaxies split by halo mass, colours, luminosity and elipticity. In Section~\ref{groups ali} we study the connection of central and group alignments. We make a further exploration on the dark matter halo distribution and central galaxy alignments in Section~\ref{dmali}, and we discuss other possible secondary dark matter halo dependences in Section~\ref{history}.
Finally, we present a summary and discuss the implications of our results in Section~\ref{discuss}. The TNG simulation adopts the standard $\Lambda$CDM cosmology from \citep{Planck2016}, with parameters $\Omega_{\rm m} = 0.3089$,  $\Omega_{\rm b} = 0.0486$, $\Omega_\Lambda = 0.6911$, $H_0 = 100\,h\, {\rm km\, s^{-1}Mpc^{-1}}$ with $h=0.6774$, $\sigma_8 = 0.8159$, and $n_s = 0.9667$.

\section{The Illustris TNG hydrodynamical simulation}
\label{sec:TNG}

Here we use the galaxy and dark-matter halo catalogues of the IllustrisTNG300-1 run at $z=0$  (hereafter TNG300, \citealt{Nelson2019}). The magneto-hydrodynamical cosmological simulations IllustrisTNG, are performed with the {\sc arepo} moving-mesh code \citep{Springel2010} and represent an updated version of the Illustris simulations \citep{Vogelsberger2014a, Vogelsberger2014b, Genel2014}. The IllustrisTNG simulations include sub-grid models that account for radiative metal-line gas cooling, star formation, chemical enrichment from SNII, SNIa and AGB stars, stellar feedback, supermassive-black-hole formation with multi-mode quasar, and kinetic black-hole feedback.

TNG300 represents the largest simulated box of the IllustrisTNG suite with the highest resolution available. 
It adopts a cubic box of $205\,h^{-1}$~Mpc side, and 
it is run with 2500$^3$ dark-matter particles of mass $4.0 \times 10^7 h^{-1} {\rm M_{\odot}}$. The initial conditions includes 2500$^3$ gas cells of mass $7.6 \times 10^6 h^{-1} {\rm M_{\odot}}$. 

A friends-of-friends (FOF) algorithm is used to identify the dark-matter haloes (also named as groups in this paper) using a linking length of 0.2 times the mean inter-particle separation \citep{Davis1985}. The substructures gravitationally bound are identified with the SUBFIND algorithm \citep{Springel2001,Dolag2009}.
Those subhaloes with a stellar component above a certain limit are defined as galaxies. We use the stellar mass of the galaxies, ${\rm M_\ast{}}$,  defined as the total mass of all stellar particles bound to each subhalo. In our analysis, we use all the simulated galaxies with stellar mass above $10^{8.5}$~M$_{\odot}$ represented by 429982 in total. We define colours in our sample using the galaxy photometric bands $g$ and $r$ provided by the IllustrisTNG database.

\subsection{Galaxy and dark matter halo properties}

Besides the galaxy properties from the IllustrisTNG database, in this work we compute the galaxy shapes. For this, we assume that galaxies are modelled by three-dimensional ellipsoids, where the axes are represented by the eigenvectors of the inertia tensor. We compute the inertia tensor by using the information of the individual stellar, dark matter particles and gas cells within the radius enclosing the half-dark matter mass of each subhalo. With this information, each element of the inertia tensor is computed as, 
\begin{equation}
    I_{i,j} = \sum_{n} m_{n} x_{i}^{n} x_{j}^{n},
\end{equation}
where $i, j$ = {1, 2, 3} correspond to the axis of the simulated box, and $m_{n}$ represents the mass of the n-th particle. The position of each individual particle is measured with respect to the centre of the subhalo it belongs, defined by using the particle with the minimum gravitational potential energy. 

We compute the three axis of each central galaxy with the eingenvalues of the inertia tensor defined as $a=\sqrt{\lambda_{a}}$,  $b=\sqrt{\lambda_{b}}$, and $c=\sqrt{\lambda_{c}}$, where $a>b>c$.  We investigate the galaxy alignments based on these three axis. Assuming that central galaxies can be described as ellipsoids (and defined analogously to eccentricity), the ellipticity of the central galaxies, $e$, is computed as,
\begin{equation}
e= \sqrt{\frac{a^2-c^2}{a^2}}.
\end{equation}

In this work we explore the role of other secondary galaxy properties on galaxy alignments such as the major mergers of central galaxies. For this, we make use the {\sc sublink} merger tree  \citep{Rodriguez-Gomez2015} available on the IllustrisTNG database, and we count the number of major mergers, $N_{mergers}$, that happen to each individual central galaxy. Following previous works, we define a major merger as those where the stellar mass ratio between galaxies is larger than 0.25 \citep{Peschken2019}.

For the dark matter haloes we use their virial mass, M$_{\rm h}$~[M$_{\odot}$], defined as the total mass enclosed within a sphere of radius R$_{\rm vir}$ enclosing a density equals to 200 times the critical density. We also study the role of the formation time of the dark matter haloes, $z_{form}$, computed as the time when it reaches half of the total mass obtained at $z=0$.
We compute the dark matter halo shapes using the same methodology as the one described above for central galaxies.

\section{Correlation Function}
\label{cf}
To statistically study the distribution of galaxies, we use the cross-correlation function, $\xi(r)$, between central galaxies of each group and different targets. This function measures the excess probability $dP$ of finding a target in the volume element $dV$ at a distance $r$ away from a central galaxy \citep{Peebles1980}:
\begin{equation*}
    dP=n\, dV(1+\xi(r))
\end{equation*}
where $n$ is the target numerical density.
Here we use two different targets: the total galaxy sample and the dark matter particles. To measure $\xi(r)$ in the simulation box, we use the \cite{Davis1983} method consisting of counting central galaxy-target pairs in a distance bin and normalising this quantity by the expected number of pairs for a homogeneous random distribution. 

We will refer to the standard $\xi(r)$, as the ``isotropic'' cross-correlation function, $\xi(r)_{iso}$, since we will later use a modified version that considers the orientation of the central object with respect to the targets (see Section~\ref{sec:anisotropic-corr}). As we have done in previous work, such as \cite{Rodriguez2022}, we performed correlation function calculations using codes developed by us for this purpose.

\begin{figure}
    \centering
	\includegraphics[width=0.99\columnwidth]{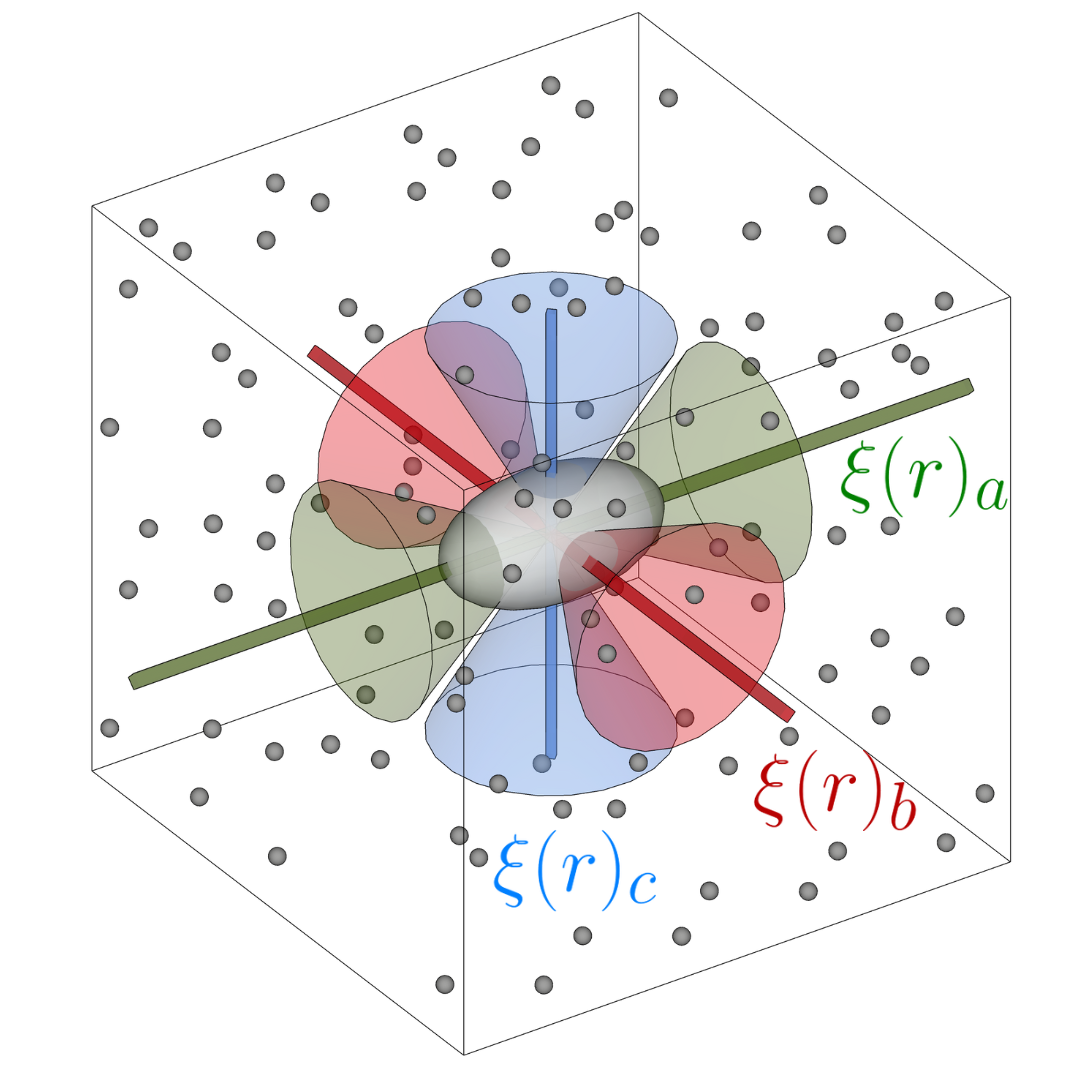}
    \caption{ Schematic showing the regions we use for the calculation of the anisotropic cross-correlation function. The green, red, and blue lines represent the major, intermediate, and minor axis of the central galaxy of the group, respectively. The shaded areas indicate the regions selected to compute the anisotropic cross-correlation for each axis.}
    \label{esquema}
\end{figure}

\subsection{Anisotropic correlation function}\label{sec:anisotropic-corr}

As mentioned above, the goal of this work is to determine whether matter tends to align with respect to the central galaxies of the groups in the TNG300 simulation, and compare the results with those from \cite{Rodriguez2022}. 
We follow the same approach as in \cite{paz2008angular} to study the galaxy alignment with respect to a particular direction.
In our implementation, we define the anisotropic cross-correlation functions as those computed using all pairs subtending an angle with respect to the shape axes smaller than a threshold ($\theta= 45^{\circ}$).
We denote $\xi(r)_{a}$ the function corresponding to the major axis, $\hat{a}$, $\xi(r)_{b}$ the one corresponding to the intermediate axis, $\hat{b}$, and $\xi(r)_{c}$ the one corresponding to the minor axis, $\hat{c}$. For a clearer understanding of this, see the schematic presented in Figure~\ref{esquema}, where green, red, and blue areas correspond to those used for $\xi(r)_{a}$, $\xi(r)_{b}$ and $\xi(r)_{c}$, respectively.  

\begin{figure}
	\includegraphics[width=\columnwidth]{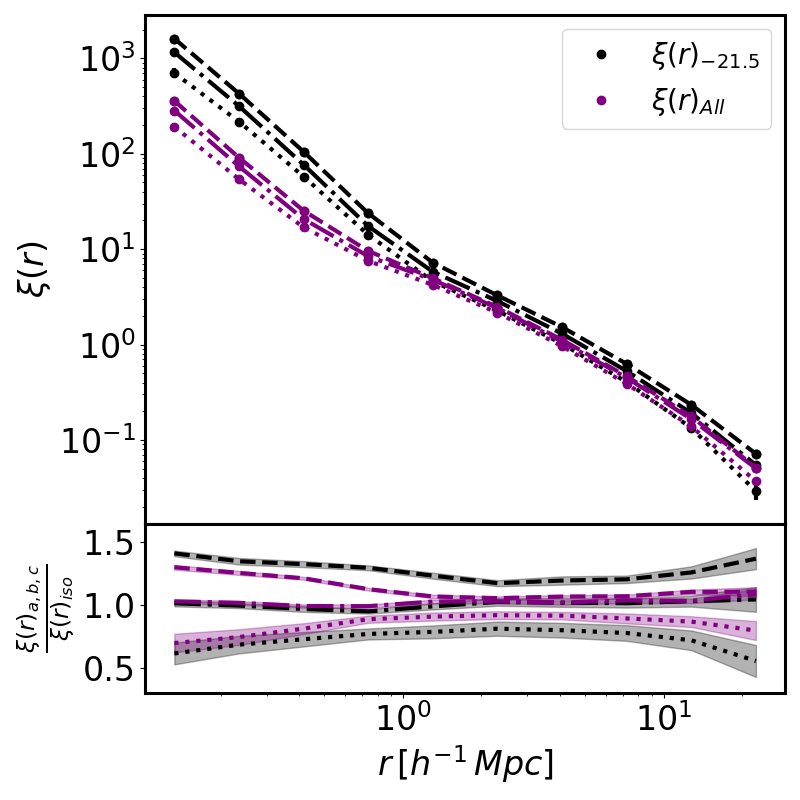}
    \caption{The anisotropic cross-correlation function using a total sample of central galaxies (violet) and those with $^{0.1}M_r<-21.5$ (black). The main panel shows the results for the $\hat{a}$ (dashed line), $\hat{b}$ (dash-dotted line) and $\hat{c}$ (dotted line) directions for both samples. The errors are computed with the jackknife method using a set of 50 subsamples and are typically smaller than the size of the points. The  subplot presents the fractional difference between the $\hat{a}$, $\hat{b}$ or $\hat{c}$ axis and the ‘isotropic’ cross-correlation functions, $\xi(r)_{iso}$. The shaded regions correspond to the error propagation.}
    \label{fig:fc1}
\end{figure}

Using the sample of galaxies described in Section~\ref{sec:TNG}, we made a first measurement of $\xi(r)_{a}$, $\xi(r)_{b}$ and $\xi(r)_{c}$  cross-correlating central galaxies with the total sample. The purple lines in the top panel of Figure~\ref{fig:fc1} represent our results.
In these measurements and throughout this paper, we use the jack-knife technique to calculate the errors of the correlation functions. For this, we split the total sample into 32 equal-size sub-boxes, representing $\sim10$~\% of the full box. 
We note, however, that most times the error bars are smaller than the points used to graph the correlation function. 
The subplot on the bottom panel of Figure~\ref{fig:fc1}  displays the ratio between the anisotropic cross-correlations ($\xi(r)_{a}$, $\xi(r)_{b}$ and $\xi(r)_{c}$) and the isotropic one ($\xi(r)_{iso}$). Hereafter, we will use this ratio to quantify the anisotropy. A ratio above or below 1 is a sign of anisotropy in the galaxy distribution.
Our results show a weak anisotropy signal for the two-halo term of the full galaxy sample.
When we restrict the sample to those halos whose central galaxy is brighter than $M_r<-21.5$ (represented by 14135 halos) the anisotropy increases (see black lines and symbols in Figure \ref{fig:fc1}). Interestingly, the two-halo term shows a significant increase in the anisotropic signal.
 Furtheremore, the correlation increases at the same time as the anisotropy. This is consistent with the results shown by \citet{Rodriguez2022} for the SDSS-DR16. Using the former sample, we analyse the dependence of the anisotropy on different properties of the galaxies and the halos in which they are located.

Our results show that the anisotropy signal is higher for the $\hat{a}$ direction than for $\hat{b}$ and $\hat{c}$, suggesting the existence of a preferential direction for the matter distribution to be aligned along the major axis. On the other hand, the anisotropic cross-correlation of the $\hat{c}$ direction is lower than the one for the $\hat{b}$. This difference indicates that the matter is mostly triaxially distributed rather than in a purely prolate configuration. This feature extends to the outskirts and is present in the 2-halo term. The distinction in the anisotropic cross-correlation function for the three galaxy axis is in agreement with \citet{paz2011alignments}, using dark matter-only simulations.

\subsection{Colour dependence}
\label{fccolordependence}

In \cite{Rodriguez2022} the authors find a strong dependence of the anisotropy with the colour of the central galaxy. Here we investigate whether the same trend is present in the TNG300. Following previous works, we define a threshold of $g-r=0.6$ to split the galaxy sample in red and blue (see Figure 1 in \citealt{Lacerna2022} for the colour distribution of the TNG300 simulated galaxies). With that threshold, 68\% of the sample are red central galaxies while 32\% are blue.

\begin{figure}
	\includegraphics[width=\columnwidth]{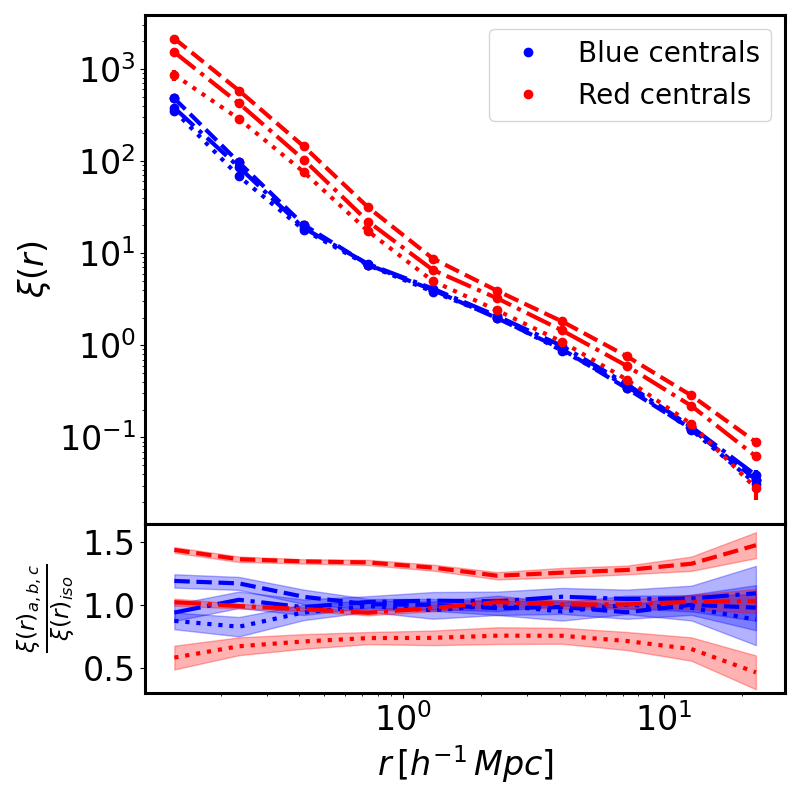}
    \caption{The anisotropic cross-correlation function in the same format as Figure \ref{fig:fc1} for the central galaxies sample with $^{0.1}M_r<-21.5$ split by colour.}
    \label{fig:fc2}
\end{figure}

Figure \ref{fig:fc2} presents the anisotropic cross-correlation functions for the groups with red and blue central galaxies. The top panel shows that the red central galaxies have a larger amplitude (i.e., are more clustered) than the blue ones. This result reflects that red central galaxies inhabit halos of higher masses.
The most important result is shown in the subplot of Figure~\ref{fig:fc2}, where we find that the red central galaxies have a significantly higher anisotropy than the blue ones. In particular, red central galaxies reach an anisotropy of $\sim$ 1.5 and 0.5 for the major and minor axis, respectively, while the intermediate axis does not show anisotropy signal.
These results suggest that the red central galaxies are aligned with the surrounding structures at least up to $\sim$ 25 Mpc. The blue central galaxies, on the other hand, have almost no anisotropy, showing only a very slight signal at small scales below $\sim 0.8 h^{-1} \, Mpc$.

These results are consistent with those obtained in \cite{Rodriguez2022}. We chose to show only these measurements because they condense the most relevant information: the red central galaxies are responsible for the anisotropy. However, when we perform cross-correlations between central galaxies and target galaxies of different colours, we get analogous results. 

\subsection{Mass dependence}
\label{fcmdependence}

\begin{figure}
	\includegraphics[width=\columnwidth]{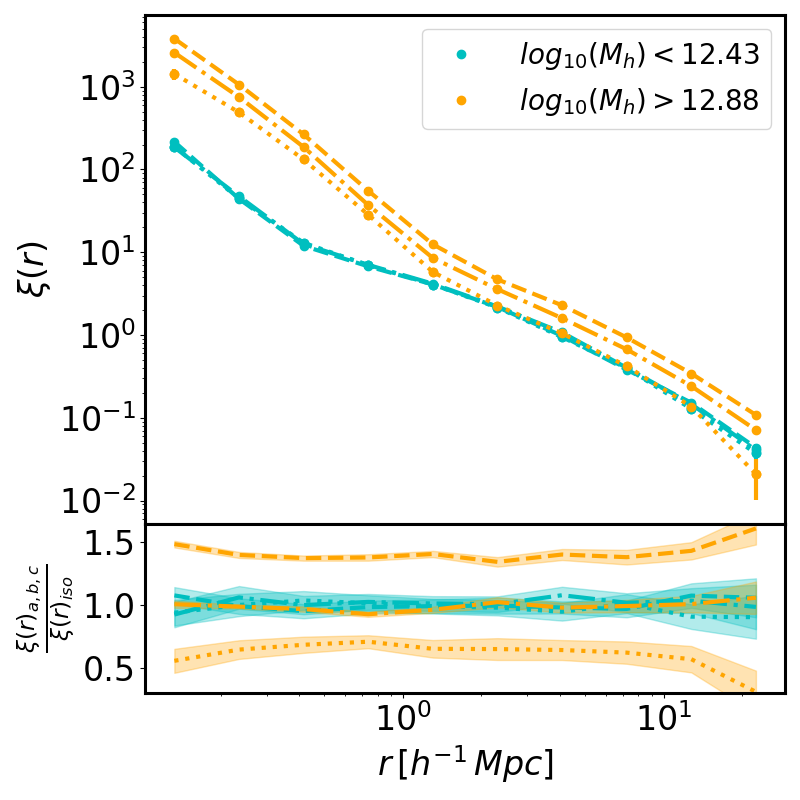}
    \caption{The anisotropic cross-correlation function in the same format as Figure \ref{fig:fc1} for the sample of central galaxies with $^{0.1}M_r<-21.5$ split by group mass.}
    \label{fig:fc3}
\end{figure}
Considering that red galaxies are responsible for the anisotropy and that they tend to inhabit more massive haloes, it would be expected a dependence of the anisotropy on the halo mass, as demonstrated by \cite{paz2011alignments}. However, a controversial result obtained in \citep{Rodriguez2022} from observations using halo mass derived from the abundance matching method was that no such dependence was detected.

Figure \ref{fig:fc3} presents the correlation functions for the first and third halo mass terciles of the distribution. For the sake of clarity, we omit the central tercile, but the corresponding curves lie between the first and third terciles.
As expected, the cross-correlation function corresponding to the more massive sample has a higher amplitude than the low-mass one (top main panel). Our findings show that massive haloes present a strong anisotropy while the less massive ones show almost no signal (see the subplot, bottom panel).
The fact that this dependence was not observed in  \cite{Rodriguez2022} could be related, in the first place, to the fact that the mass range covered was smaller  (where the three halo mass bins were defined as $\log_{10}(M_{h}/M_\odot) = [<13.43, 13.43-13.65 >13.65]$).
In addition, it is natural that the masses have more observational uncertainties due to several factors, such as the abundance matching method applied for their determination, errors in the identification of galaxy systems, or the assignment of the central galaxy.

Our results are consistent with previous findings from other cosmological simulations \citep[see][]{Shao2016,Welker2018,Tenneti2021}, suggesting that the trends are robust irrespective of sub-grid physics in the simulation. In particular \citet{Tenneti2021} show that the alignment angle between satellite galaxies and the major axis of the central galaxy strongly depends on the dark matter halo, and similar results are found for IllustrisTNG and MassiveBlack-II.

\subsection{Ellipticity dependence}
\label{fcedependence}

\begin{figure}
	\includegraphics[width=\columnwidth]{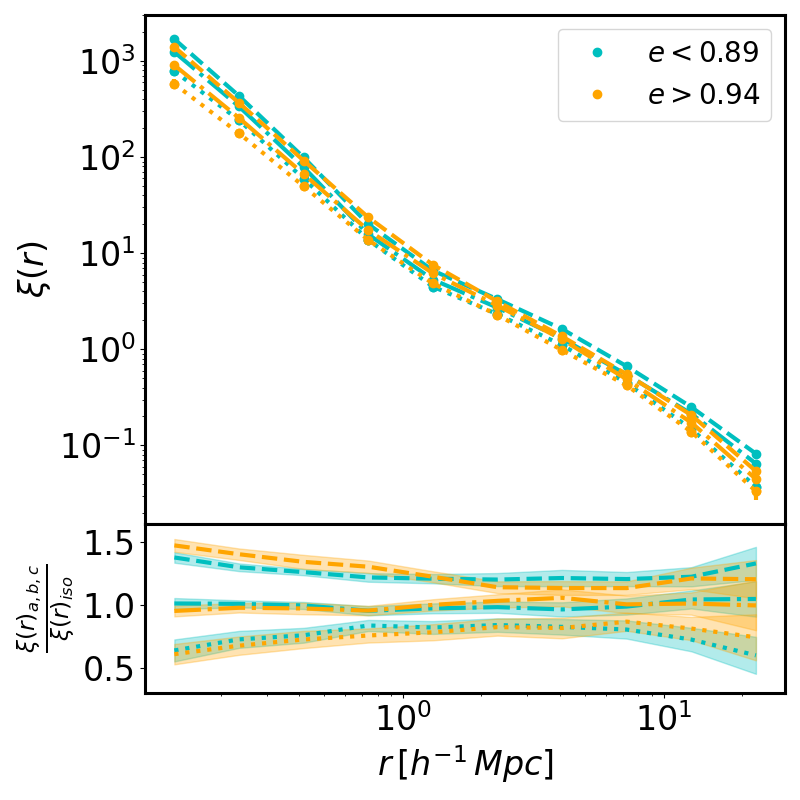}
    \caption{The anisotropic cross-correlation function in the same format as Figure \ref{fig:fc1} for the central galaxies with $^{0.1}M_r<-21.5$ split by group ellipticity. In the figure, we show the first and third terciles of for the group ellipticity.}
    \label{fig:fc4}
\end{figure}
We now study the dependence of the anisotropy on the ellipticity of the central galaxies determined from the eigenvectors of the shape tensor calculated using the stellar particles within two radii at half mass.

Figure \ref{fig:fc4}  shows the cross-correlation functions and anisotropies determined for the first and third terciles of the ellipticity distribution for the central galaxies. No differences are observed either in the correlation function or in the anisotropy. These findings support the results found in \cite{Rodriguez2022}. This is relevant because, in the observations, the ellipticity is affected by projection effects and the anisotropy signal could be distorted. The simulation is free of these projection effects, yet no dependence of anisotropy on ellipticity is observed.

\section{Central and group alignment}
\label{groups ali}

As a final comparison with the observational results of \cite{Rodriguez2022}, we propose to determine the anisotropic correlation functions but, in this case, relative to the direction of the group axis. We calculate these directions from the eigenvectors associated with the eigenvalues of the shape tensor:
\begin{equation}
\label{eq1}
    I_{ij}= \frac{1}{N_g}\sum_{\alpha=1}^{N_g} X_{\alpha i} X_{\alpha j},
\end{equation}
where $X_{\alpha i}$ is the $i^{th}$ component of the Cartesian coordinates of the $\alpha$ object relative to the centre of the group and $N_{g}$ is the number of objects in the group.

\begin{figure}
	\includegraphics[width=\columnwidth]{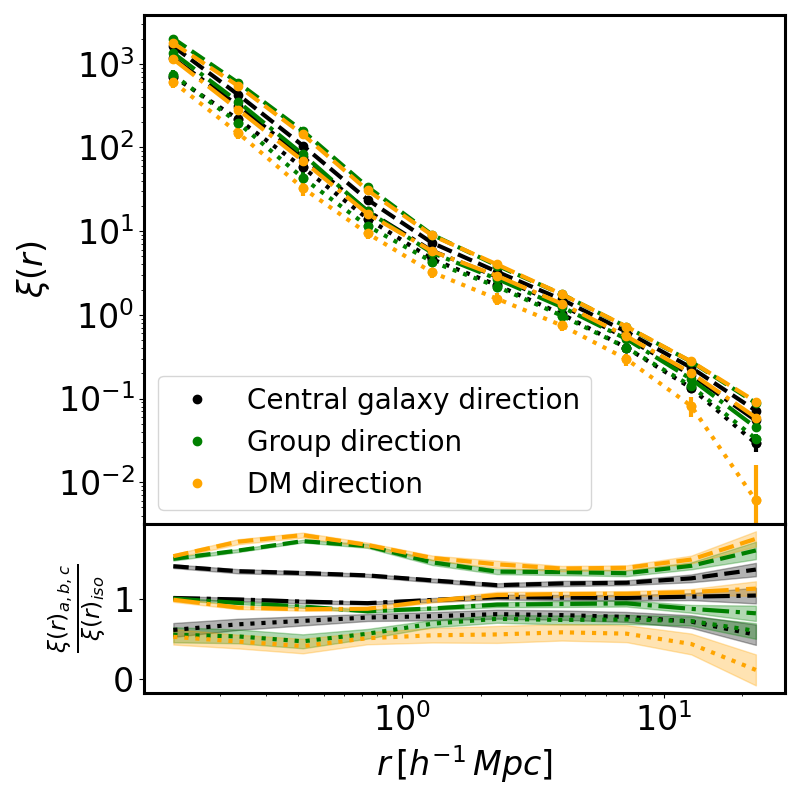}
    \caption{
    Comparison between the anisotropic cross-correlation function using three different shape estimators. Black lines represent the sample of central galaxies with $^{0.1}M_r<-21.5$ in the directions as shown in Figure \ref{fig:fc1} (black), green lines  the main axis of the shape of the galaxy group in which they reside (green), and calculated with dark matter particles (orange).
    We present our results for the three directions $\hat{a}$, $\hat{b}$ or $\hat{c}$.
    }
    \label{fig:fc5}
\end{figure}

Figure \ref{fig:fc5} presents the comparison between the result obtained when using the axis from the central galaxies shape (presented in Figure \ref{fig:fc1}, black lines) and the one using the axis from the groups shape (green lines). As in \citet{Rodriguez2022}, we find that the one-halo term alignment signal is larger for the case when the group shape is used.

However, the difference found in TNG300 is smaller than the result obtained from observations. This can be explained by projection effects, together with the method used to calculate the correlation function, both enhancing the anisotropies on the smaller scales. 
A striking result obtained from the observations was that both the anisotropy associated with the central galaxy and the group were indistinguishable on large scales. However, when we analyze it in the simulation, we can detect a small difference, which means that the groups are more aligned with the distant environment than their central galaxy. This result suggests that the observed anisotropy signal for the central galaxy could be because of an alignment between it and its group. 
It is possible that two effects are involved in this phenomenon. First, non-linear physical processes would produce the alignment of the central galaxy within the dark matter halo. Second, the large-scale structure distribution of matter is imprinted in the hierarchical nature of the dark matter halo formation \citep{paz2011alignments}. This mechanism would explain the galaxy cluster alignment at large scales beyond the halo.

Taking advantage of the dark matter information in the simulation, we calculate the halo shape with the same procedure as in equation \ref{eq1} but using the dark matter particles. The results with the latter determination are shown in orange in Figure \ref{fig:fc5}. We find a similar trend to the results obtained when member galaxies are used to estimate the group shapes, but the alignment increases slightly. Our results are at some point expected since over-dense dark matter regions are linked with the  satellite galaxies' position within the halos.

\begin{figure}
	\includegraphics[width=\columnwidth]{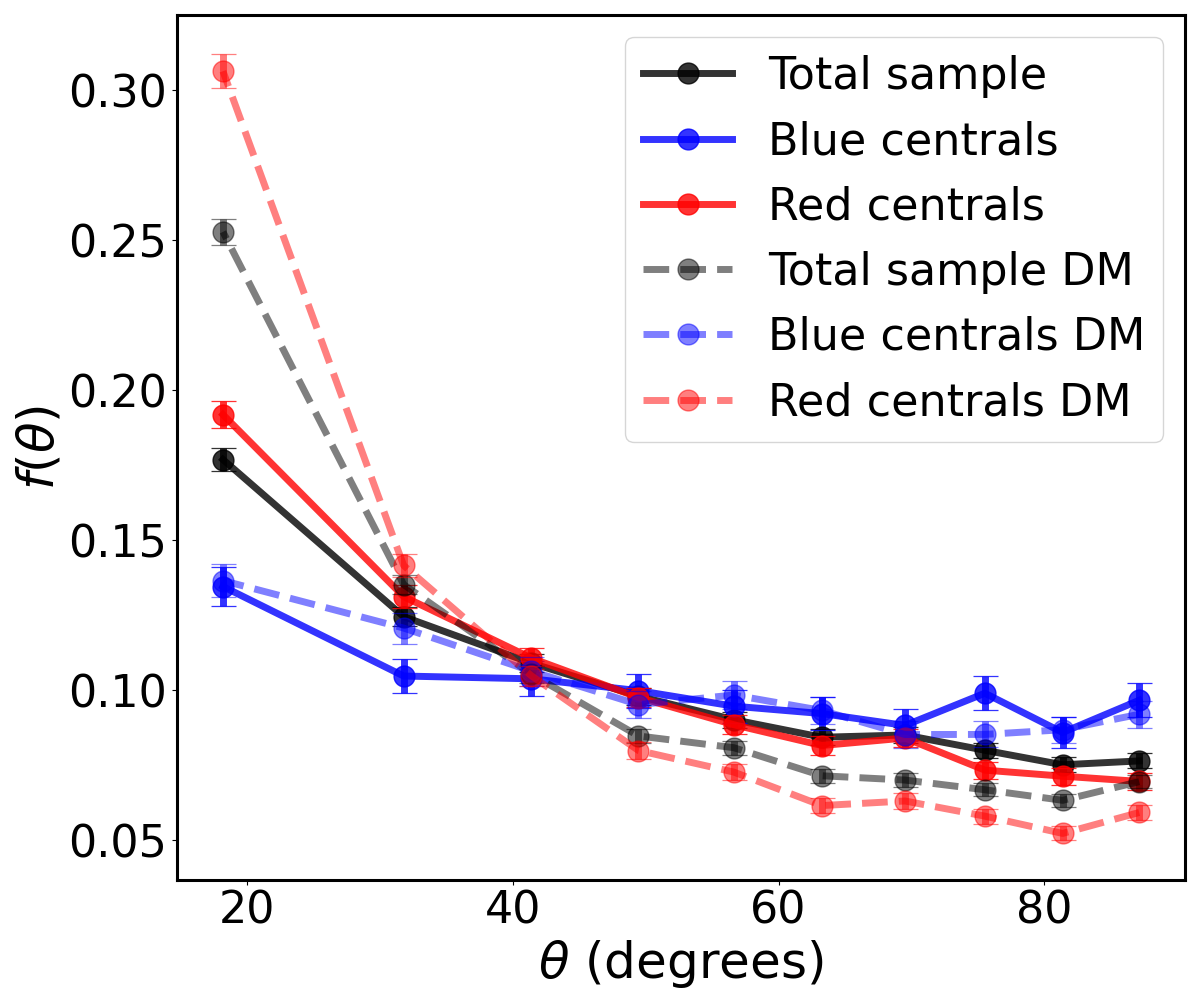}
    \caption{The measured probability density distribution of the alignment between the central galaxy and the group shape for the total sample of central galaxies with $^{0.1}M_r<-21.5$, and separating these central galaxies by colour. Solid lines correspond to the halo shape determinations using the galaxies residing in the halo, while dashed lines are the measurements with the dark matter particles.}
    \label{fig:ali1}
\end{figure}

Since the alignment obtained by calculating the shape of the halo with galaxies or dark matter is very similar and slightly larger than that estimated with the central galaxy shape, it is clear that the alignment between the central galaxy and its host group/halo needs to be explored.
To quantify this alignment, we measure the angle between the corresponding major axes. Figure \ref{fig:ali1} shows the measured probability density distribution for total, red, and blue samples of central galaxies regarding the halo shape calculated with the galaxies members (solid lines) and with the dark matter particles (dashed line). 
As expected from Figure \ref{fig:fc5}, the alignment of the central galaxy with the dark matter in its host halo is slightly stronger. 

Therefore, in both cases, the red galaxies show more alignments than the total while the blue galaxies show a very low alignment signal. 
As in \cite{Rodriguez2022}, when we divide the sample into red and blue central galaxies, we obtain a larger alignment signal for the red galaxies while it almost disappears for the blue ones. Looking at the alignment calculated using dark matter, it can be seen that while the alignment for the red central galaxies increases, it remains the same for the blue ones.
To better understand this result, in the next section we will concentrate on dark matter.

\section{Dark Matter alignment}
\label{dmali}
\begin{figure}
	\includegraphics[width=\columnwidth]{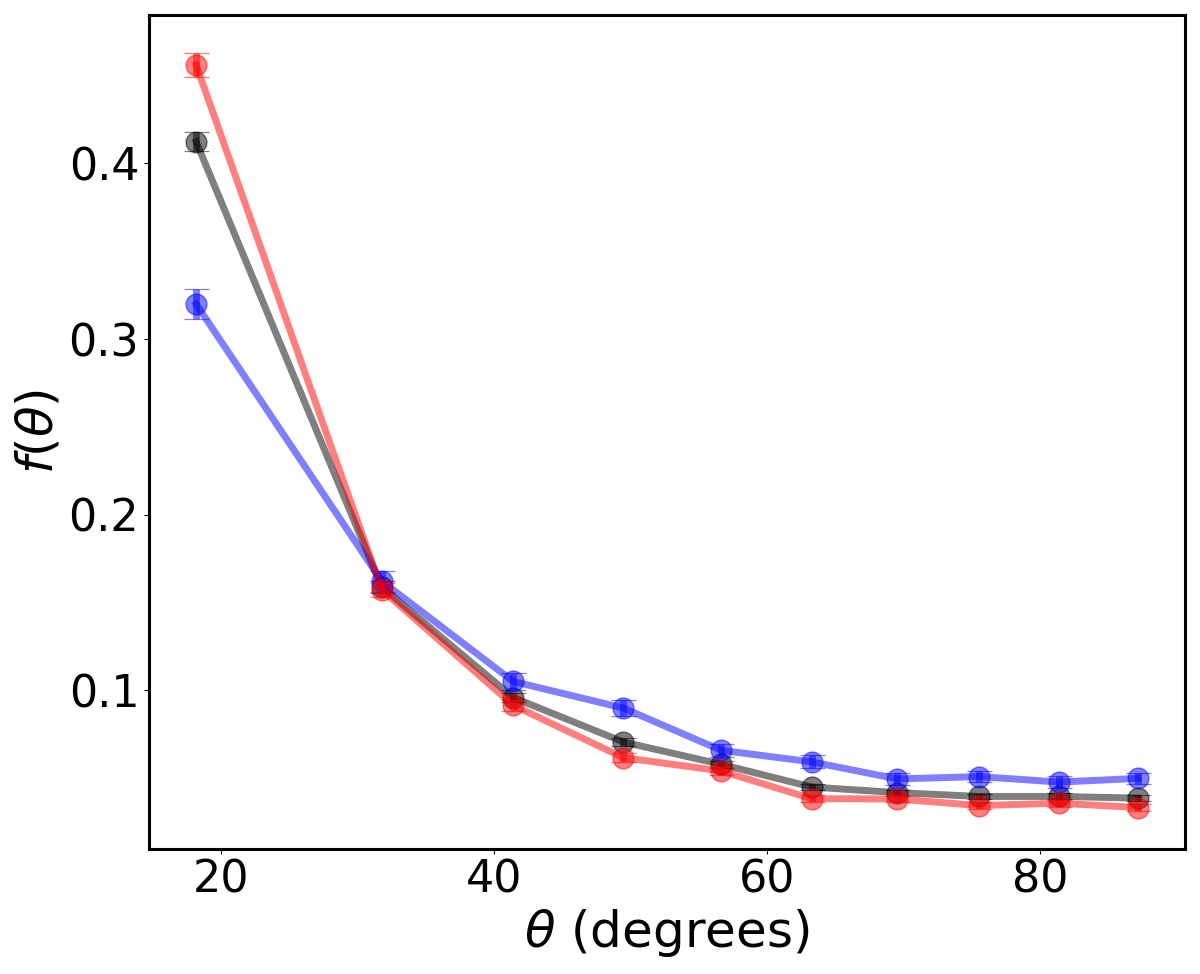}
    \caption{The measured probability density distribution of the alignment between the dark matter of the central galaxy and the dark matter of the group shape for the same samples as Figure \ref{fig:ali1}.}
    \label{fig:alidmdm}
\end{figure}

In the previous section, we show that the stellar component of blue galaxies is not aligned with the dark matter of the host halo. At this stage, we want to see if the dark matter in the central galaxies behaves similarly.
In Figure \ref{fig:alidmdm}, we show the alignment between the main axis of the central galaxy and the corresponding to the host halo, both calculated using the dark matter particles and following the same procedure as in Section \ref{groups ali}. 
As can be seen, both the total sample and red galaxies maintain a similar behavior to that found when the stellar component was used to calculate their shapes. Only a slight increase in alignment is observed. However, something very different happens with the blue galaxies, which in this case, have a considerably higher alignment than the one found in figure \ref{fig:ali1}.

\begin{figure}
	\includegraphics[width=\columnwidth]{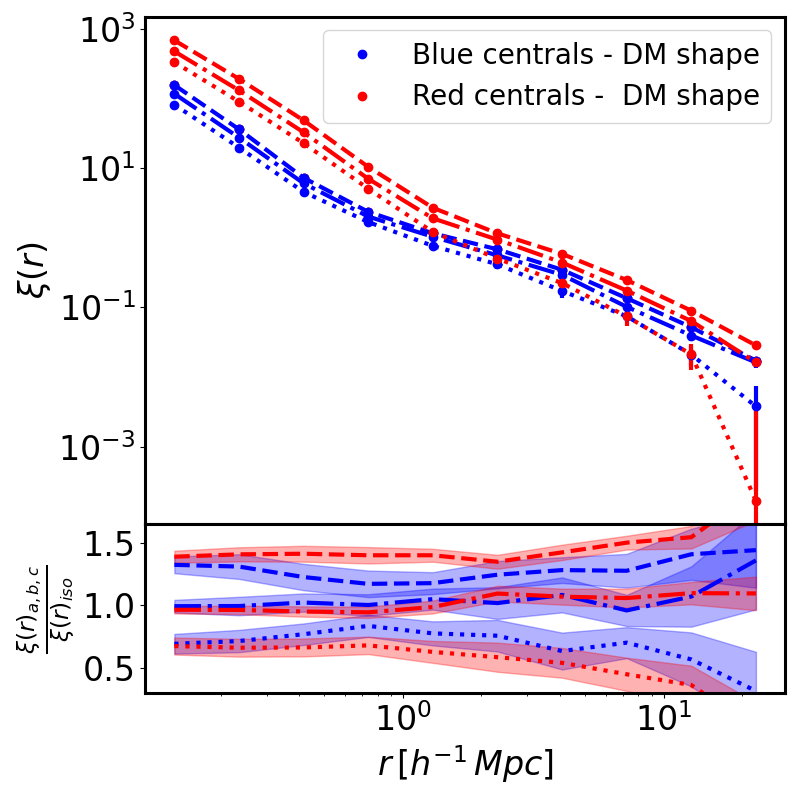}
    \caption{The anisotropic cross-correlation function in the same format as Figure \ref{fig:fc2} but taking as axes those obtained from the shape of the dark matter in the sub-halo containing the central galaxy.}
    \label{fig:fc6}
\end{figure}

The above result should be evidenced in the correlation function if we use dark matter instead of stars.
Figure \ref{fig:fc6} shows the correlation function and anisotropy resulting from considering the central galaxy with the three shape axes calculated from the dark matter particles of the subhalo it inhabits and using dark matter particles as targets.
In contrast to what we obtained in Figure \ref{fig:fc2} both the central blue and red galaxies show a strong anisotropy signal. Although the anisotropy of the blue central galaxies is slightly lower, it is comparable to that of the red ones, both evidently aligned with the host halo and surrounding structure.

\begin{figure}
	\includegraphics[width=\columnwidth]{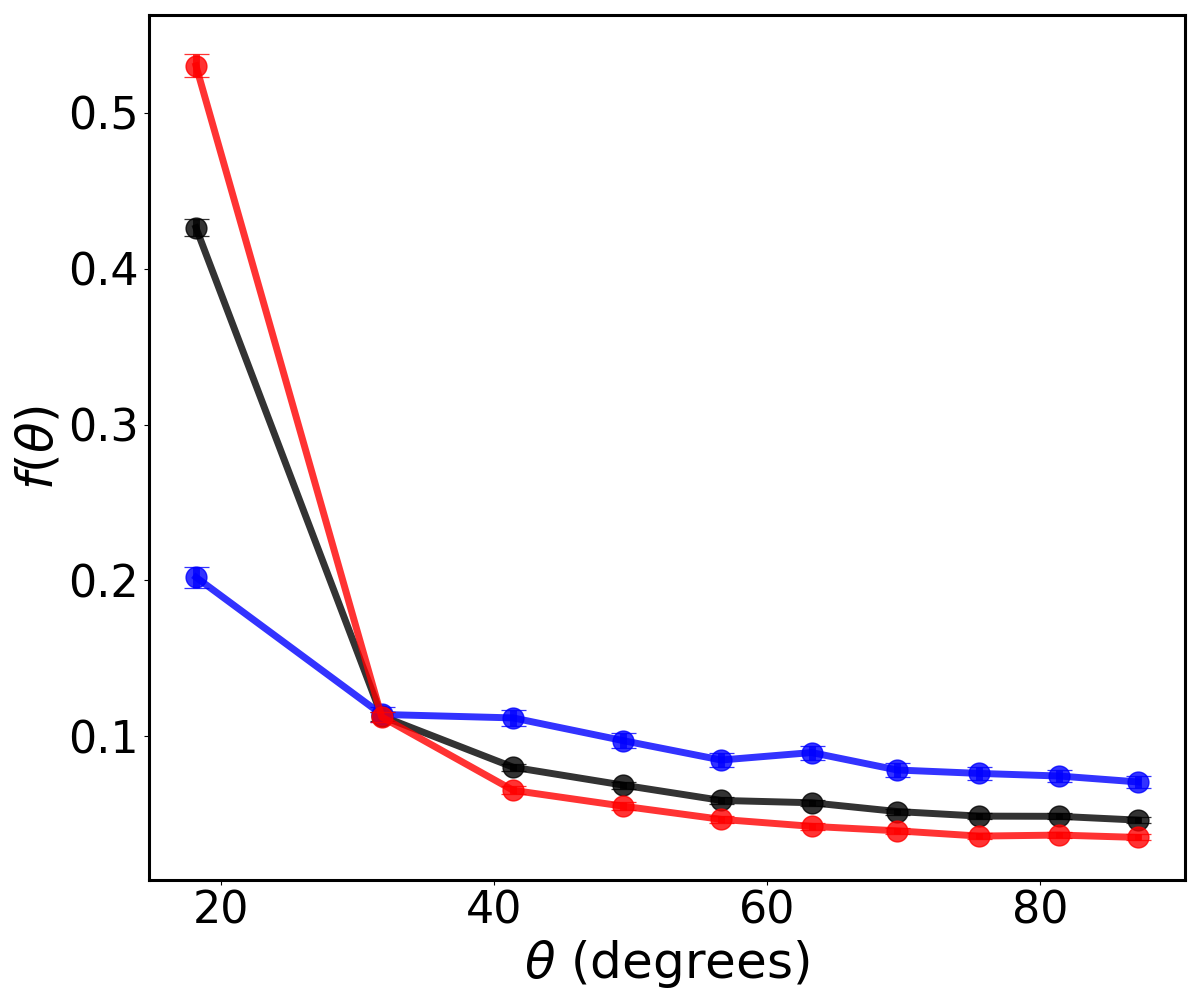}
    \caption{The measured probability density distribution of the alignment between the dark matter and stars on the central galaxy for the same samples as Figure \ref{fig:ali1}.}
    \label{fig:alistardm}
\end{figure}

These latest results suggest a misalignment between the stellar and dark matter components for blue galaxies. Consequently, we study the alignment between stars and dark matter in the sub-halo inhabited by the central galaxies. We present this in Figure \ref{fig:alistardm}. Evidently, the main axis of the red central galaxies is strongly aligned with that of their halo, while that of the blue ones is close to a random distribution.
That is, the stars are good tracers of the shape of the subhalo of the red central galaxies but not of the blue ones. This allows us to understand why this alignment occurs up to scales larger than $\sim$ 10 Mpc: The stars of the red central galaxies are aligned with their dark matter subhalo, which is aligned with the halo containing the group, which, in turn, is aligned with the surrounding structures. The behavior of blue central galaxies could be explained in the same way, but, unlike the red ones, they have the stellar component misaligned with their own subhalo, which makes the anisotropy signal fade when only stars are considered. 

\section{Secondary dependencies of anisotropies for red galaxies }
\label{history}

\begin{figure}
\includegraphics[width=0.89\columnwidth]{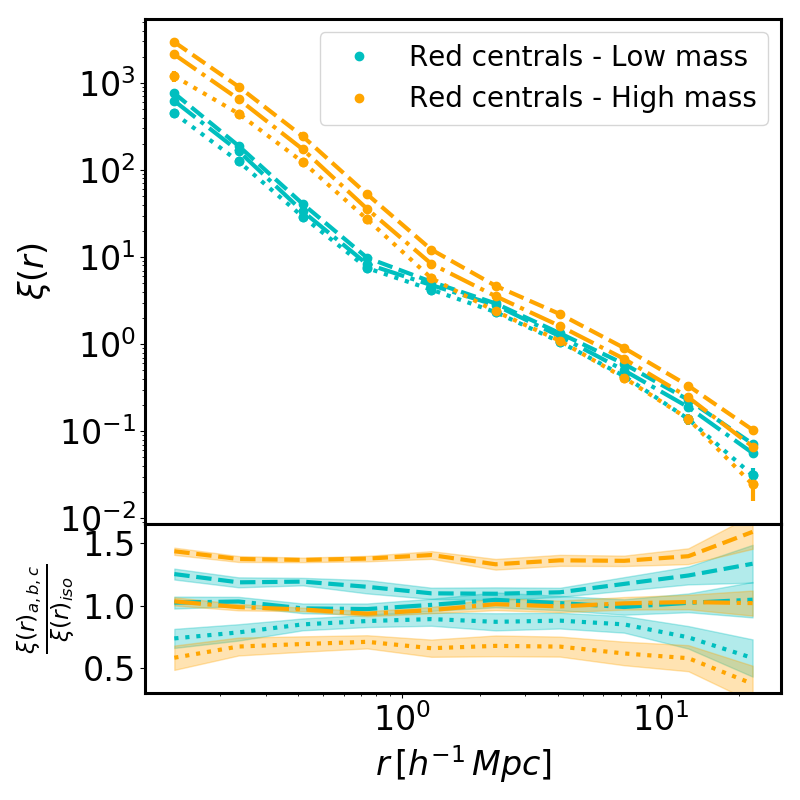}
	\includegraphics[width=0.89\columnwidth]{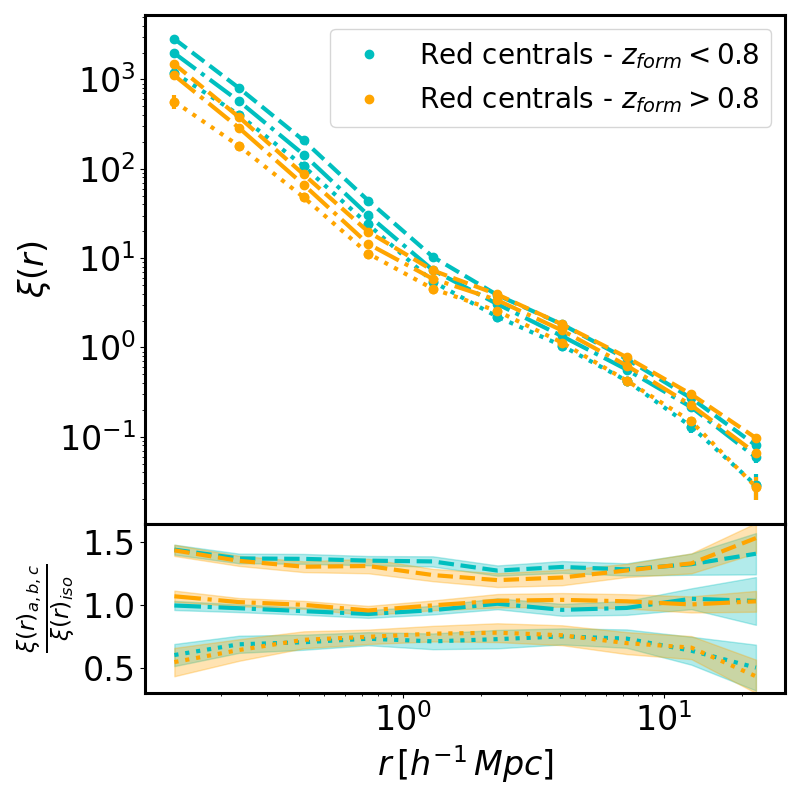}
	\includegraphics[width=0.89\columnwidth]{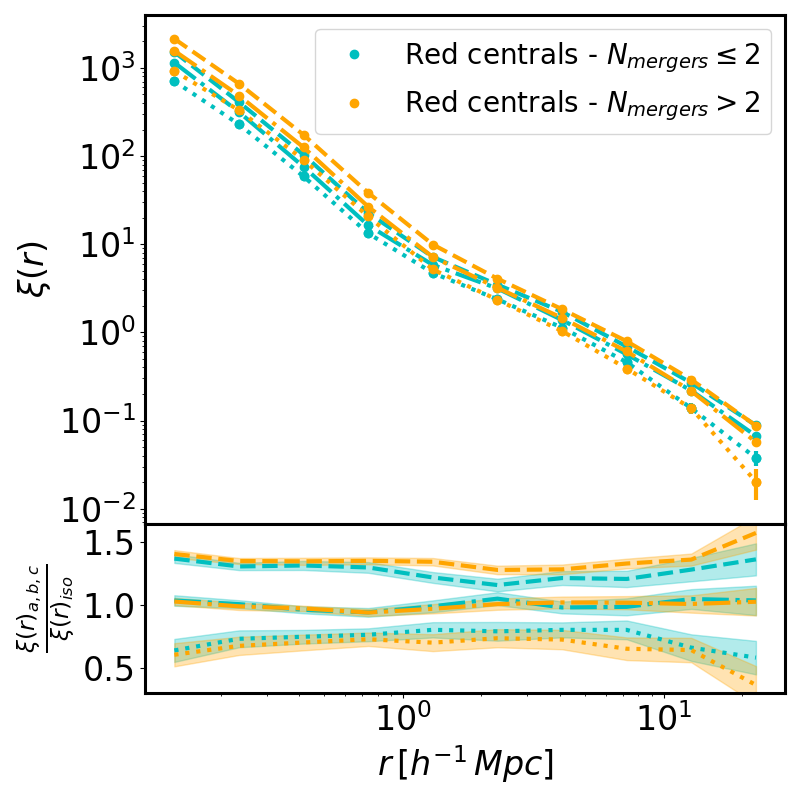}
    \caption{The anisotropic cross-correlation function for red central galaxies to analize dependence of halo mass (top panel), formation redshift (central panel) and the number of major mergers (bottom panel).
   }
    \label{fig:secondary}
\end{figure}

In previous sections, we show that red central galaxies are aligned with their surroundings, while the blue ones are misaligned. In addition, we find a dependence of the alignment of the central galaxies on the halo mass they inhabit. 
This leads us to think that there may be other (secondary) dependencies linked to the evolutionary processes of galaxy assembly and mass accretion history. Two properties that hold information of this type are the redshift at which half of the haloes' mass was assembled ($z_{form}$) and the number of major mergers (of one-fifth of their mass) they underwent during their lifetime ($N_{mergers}$).

In previous sections we show that blue central galaxies do not present any alignment signal when we split the sample by halo mass regardless of the parameters used and the selection within the sample. Consequently, in this section we focus on the results corresponding to red central galaxies. 

We first split the red central galaxies according to the mass of the halo they inhabit. We divide the sample by the median of the mass distribution, whose value is $\log _{10} (M_h)=12.86$. Then, following section \ref{cf}, we calculate the cross-correlation function and the anisotropy for high-mass and low-mass samples. We present the results for these samples in the top panel of Figure \ref{fig:secondary}. Red central galaxies hosted in massive haloes are more aligned with the surrounding structures than those in less massive haloes. This means that the dependence of anisotropy on mass is still present even when we only consider red central galaxies. This outcome supports what we found in the previous section: the anisotropy signal on large scales is more related to the host halo than the central galaxy. It is pertinent to mention that, as the stellar mass of the central galaxy correlates very well with the host halo mass, we reach similar results if we perform the above analysis, but in terms of this mass instead of that of the halo.

We will now analyse the dependence of anisotropy on $z_{form}$.
Following the same procedure as in the previous paragraphs, we split the sample according to the median of the $z_{form}$ distribution. We present the results in the middle panel of Figure \ref{fig:secondary}. 
Although the correlation function shows some difference in the one halo term, the variation in anisotropy for halos with different $z_{form}$ is negligible. 

We also investigate if the number of major mergers, $N_{mergers}$, has an impact on the anisotropy. 
In the lower panel of Figure \ref{fig:secondary}, we present the results corresponding to the samples with a number of mergers above and below the median value ($N_{mergers}=2$). 
As in the formation redshift study, we observe minor differences in the one halo term of the correlation function. However, we did not find relevant differences in the anisotropies. We carry out this analysis considering the number of major mergers based on the stellar material, but the results are similar when we use the dark matter mass budget to define the mergers.

Unexpectedly, we do not find a dependence on the redshift and the number of mergers. However, we know from Figures \ref{fig:fc3} and \ref{fig:secondary} that there is a strong dependence on the mass. Consequently, we investigated the influence of the redshift and the number of mergers on the anisotropy for the highest mass bin. This selection yielded no difference in the anisotropic correlation functions, suggesting that the colour and halo masses may be the properties that best summarise information about anisotropy. We do not include figures for these tests because they are very similar to those already presented in Figure \ref{fig:secondary}. 

\section{Discussion and conclusions }
\label{discuss}

We use the cosmological hydrodynamical simulation IllustrisTNG to determine the alignment of the central galaxies with the environment to deepen the observational results obtained by \cite{Rodriguez2022}.

When we restrict the sample by the colour or brightness of the central galaxies, our results show good agreement with the results obtained in \cite{Rodriguez2022}, i.e. the brighter central galaxies show more anisotropy and, within this subsample, the red central galaxies are responsible for most of the signal.

When we consider more specific galaxy properties, such as the ellipticity of the central galaxy and the mass of the host halo, we find a dependence of the anisotropy on the mass but not on the ellipticity. While the result for the ellipticity agrees with the observations, this does not happen for the mass, where the observations do not show any dependence. This discrepancy is probably because of the way the masses of the galaxy groups are assigned in the observations. Although the abundance matching method used to estimate the masses of galaxy groups is statistically better than dynamical estimates, it is still not sufficiently accurate for the determination of individual masses. This would explain the observational inability to detect the dependence of the anisotropy on the mass observed in the simulations. 
The fact that, in observations, it is possible to detect the dependence of the anisotropy on the colour of the central galaxy, but not on the mass of the host halo, would show that the colour of the central galaxy is a better indicator of the mass of the host halo than the masses assigned by traditional methods. 

A remarkable result of this work is the detected increase of the anisotropy with the host halo mass taking into account the shape of the central galaxy. According to the determinations of other authors, the dependence of the anisotropy with the mass associated with the shape of the host halo is well known. So the alignment of the central galaxy with the environment on large scales could be an effect of the alignment of the central galaxy and its host halo.

From the comparison of the anisotropies estimated using the shapes of central galaxies or groups, we observe that the signal of the one-halo term is higher in the groups, while they are similar in the two-halo term. This supports the reasoning in the previous paragraph that the alignment of the central galaxy with the large-scale structure is a consequence of the alignment with the halo it inhabits. To figure this out, we studied the alignment between the central galaxy and the host halo. We find that, in general, there is a remarkable alignment, with the red central galaxies being most closely aligned with their host halo. This result is in agreement with previous observational results.

All the above analysis was performed using the stellar component of the simulation. However, since various astrophysical processes could influence the orientation of the central galaxy, we study the behavior of the anisotropy based on the shapes determined using dark matter particles. Using these determinations, we repeat the analysis of the previous paragraph and observe that, remarkably, the alignment of both red and blue central galaxies increases. When we study the anisotropy using the correlation function, we find that unlike the case when we calculated the shape with the stars, the blue central galaxies present a similar anisotropy signal to the red galaxies. This leads us to understand the origin of the lack of anisotropy in the blue central galaxies. Analysing the alignment between stars and dark matter in the central galaxies, we find that the blue galaxies are misaligned with their own halo compared to the red galaxies, which show a significant alignment.

Since the alignment is dominated by the colour of the central galaxy in both observations and simulations, it is worth wondering whether the alignment is actually determined by the assembly process of the central galaxy. To see this, we study for red central galaxies (because they show an anisotropy signal) the dependence of the anisotropy on the mass of the host halo, the formation redshift and the number of mergers. While the dependence on the host halo mass resembles that of the total sample, unexpectedly, the anisotropy seems to be neither sensitive to the formation redshift nor the number of mergers.

The results found in this work can be summarized as follows: the alignment signal or anisotropy found from the shape axes of the central galaxies is the consequence of a concatenation of alignments. Starting at the smallest scales, the baryonic material of the central galaxy aligns with the dark matter of its own halo, which in turn aligns with the host halo, and eventually, the host halo aligns with the structures surrounding it. The unexpected independence of the alignment signal with respect to the formation redshift and the number of mergers does not allow us to link the assembly process of the central galaxy with the detected anisotropy. It is probably necessary to study in detail and individually the alignment process of the central galaxies both with their own halo and with the host halo, to explain the processes that determine the alignment or misalignment of the baryonic material with the halos of dark matter. Other halo properties such as the tidal field and the evolution of the galaxy alignment will be evaluated in the future for further exploration. This is essential to understand observations in which only the two ends of this sequence can be measured.

\section*{Acknowledgements}
The authors wish to thank the anonymous referee for his/her report that helps us to improve this manuscript.
FR and MM thanks the support by Agencia Nacional de Promoci\'on Cient\'ifica y Tecnol\'ogica, the Consejo Nacional de Investigaciones Cient\'{\i}ficas y T\'ecnicas (CONICET, Argentina) and the Secretar\'{\i}a de Ciencia y Tecnolog\'{\i}a de la Universidad Nacional de C\'ordoba (SeCyT-UNC, Argentina).
MCA acknowledges financial support from the Seal of Excellence @UNIPD 2020 program under the ACROGAL project.
FR would like to thank Rocío Rodríguez and Santiago Díaz for their help in producing a graphic to make our work clearer. 
\section*{Data Availability}
 The data underlying this article will be shared on reasonable request to the corresponding authors. 


\bibliographystyle{mnras}
\bibliography{main} 

\begin{thebibliography}{}
\makeatletter
\relax
\def\mn@urlcharsother{\let\do\@makeother \do\$\do\&\do\#\do\^\do\_\do\%\do\~}
\def\mn@doi{\begingroup\mn@urlcharsother \@ifnextchar [ {\mn@doi@}
  {\mn@doi@[]}}
\def\mn@doi@[#1]#2{\def\@tempa{#1}\ifx\@tempa\@empty \href
  {http://dx.doi.org/#2} {doi:#2}\else \href {http://dx.doi.org/#2} {#1}\fi
  \endgroup}
\def\mn@eprint#1#2{\mn@eprint@#1:#2::\@nil}
\def\mn@eprint@arXiv#1{\href {http://arxiv.org/abs/#1} {{\tt arXiv:#1}}}
\def\mn@eprint@dblp#1{\href {http://dblp.uni-trier.de/rec/bibtex/#1.xml}
  {dblp:#1}}
\def\mn@eprint@#1:#2:#3:#4\@nil{\def\@tempa {#1}\def\@tempb {#2}\def\@tempc
  {#3}\ifx \@tempc \@empty \let \@tempc \@tempb \let \@tempb \@tempa \fi \ifx
  \@tempb \@empty \def\@tempb {arXiv}\fi \@ifundefined
  {mn@eprint@\@tempb}{\@tempb:\@tempc}{\expandafter \expandafter \csname
  mn@eprint@\@tempb\endcsname \expandafter{\@tempc}}}

\bibitem[\protect\citeauthoryear{{Agustsson} \& {Brainerd}}{{Agustsson} \&
  {Brainerd}}{2010}]{Agustsson2010}
{Agustsson} I.,  {Brainerd} T.~G.,  2010, \mn@doi [\apj]
  {10.1088/0004-637X/709/2/1321}, \href
  {https://ui.adsabs.harvard.edu/abs/2010ApJ...709.1321A} {709, 1321}

\bibitem[\protect\citeauthoryear{Ahumada et~al.,}{Ahumada
  et~al.}{2020}]{ahumada2020}
Ahumada R.,  et~al., 2020, The Astrophysical Journal Supplement Series, \href
  {https://iopscience.iop.org/article/10.3847/1538-4365/ab929e"} {249, 3}

\bibitem[\protect\citeauthoryear{{Bray} et~al.,}{{Bray}
  et~al.}{2016}]{Bray2016}
{Bray} A.~D.,  et~al., 2016, \mn@doi [\mnras] {10.1093/mnras/stv2316}, \href
  {https://ui.adsabs.harvard.edu/abs/2016MNRAS.455..185B} {455, 185}

\bibitem[\protect\citeauthoryear{Catelan, Kamionkowski  \& Blandford}{Catelan
  et~al.}{2001}]{Catelan2001}
Catelan P.,  Kamionkowski M.,   Blandford R.~D.,  2001, Monthly Notices of the
  Royal Astronomical Society, \href
  {https://doi.org/10.1046/j.1365-8711.2001.04105.x} {320, L7}

\bibitem[\protect\citeauthoryear{Ciotti \& Dutta}{Ciotti \&
  Dutta}{1994}]{Ciotti1994}
Ciotti L.,  Dutta S.,  1994, Monthly Notices of the Royal Astronomical Society,
  \href {https://doi.org/10.1093/mnras/270.2.390} {270, 390}

\bibitem[\protect\citeauthoryear{Crittenden, Natarajan, Pen  \&
  Theuns}{Crittenden et~al.}{2001}]{Crittenden2001}
Crittenden R.~G.,  Natarajan P.,  Pen U.-L.,   Theuns T.,  2001, The
  Astrophysical Journal, \href {https://doi.org/10.1086/322370} {559, 552}

\bibitem[\protect\citeauthoryear{Davis \& Peebles}{Davis \&
  Peebles}{1983}]{Davis1983}
Davis M.,  Peebles P.,  1983, The Astrophysical Journal, 267, 465

\bibitem[\protect\citeauthoryear{{Davis}, {Efstathiou}, {Frenk}  \&
  {White}}{{Davis} et~al.}{1985}]{Davis1985}
{Davis} M.,  {Efstathiou} G.,  {Frenk} C.~S.,   {White} S.~D.~M.,  1985,
  \mn@doi [\apj] {10.1086/163168}, \href
  {http://adsabs.harvard.edu/abs/1985ApJ...292..371D} {292, 371}

\bibitem[\protect\citeauthoryear{{Deason} et~al.,}{{Deason}
  et~al.}{2011}]{Deason2011}
{Deason} A.~J.,  et~al., 2011, \mn@doi [\mnras]
  {10.1111/j.1365-2966.2011.18884.x}, \href
  {https://ui.adsabs.harvard.edu/abs/2011MNRAS.415.2607D} {415, 2607}

\bibitem[\protect\citeauthoryear{{Dolag}, {Borgani}, {Murante}  \&
  {Springel}}{{Dolag} et~al.}{2009}]{Dolag2009}
{Dolag} K.,  {Borgani} S.,  {Murante} G.,   {Springel} V.,  2009, \mn@doi
  [\mnras] {10.1111/j.1365-2966.2009.15034.x}, \href
  {https://ui.adsabs.harvard.edu/abs/2009MNRAS.399..497D} {399, 497}

\bibitem[\protect\citeauthoryear{{Dong}, {Lin}, {Kang}, {Ocean Wang}, {Dutton}
  \& {Macci{\`o}}}{{Dong} et~al.}{2014}]{Dong2014}
{Dong} X.~C.,  {Lin} W.~P.,  {Kang} X.,  {Ocean Wang} Y.,  {Dutton} A.~A.,
  {Macci{\`o}} A.~V.,  2014, \mn@doi [\apjl] {10.1088/2041-8205/791/2/L33},
  \href {https://ui.adsabs.harvard.edu/abs/2014ApJ...791L..33D} {791, L33}

\bibitem[\protect\citeauthoryear{Fleck \& Kuhn}{Fleck \&
  Kuhn}{2003}]{Fleck2003}
Fleck J.-J.,  Kuhn J.~R.,  2003, The Astrophysical Journal, \href
  {https://doi.org/10.1086/375585} {592, 147}

\bibitem[\protect\citeauthoryear{{Genel} et~al.,}{{Genel}
  et~al.}{2014}]{Genel2014}
{Genel} S.,  et~al., 2014, \mn@doi [\mnras] {10.1093/mnras/stu1654}, \href
  {https://ui.adsabs.harvard.edu/abs/2014MNRAS.445..175G} {445, 175}

\bibitem[\protect\citeauthoryear{{Jagvaral}, {Singh}  \&
  {Mandelbaum}}{{Jagvaral} et~al.}{2022}]{Jagvaral2022}
{Jagvaral} Y.,  {Singh} S.,   {Mandelbaum} R.,  2022, \mn@doi [\mnras]
  {10.1093/mnras/stac1424}, \href
  {https://ui.adsabs.harvard.edu/abs/2022MNRAS.514.1021J} {514, 1021}

\bibitem[\protect\citeauthoryear{Jing}{Jing}{2002}]{Jing2002}
Jing Y.,  2002, Monthly Notices of the Royal Astronomical Society, \href
  {https://doi.org/10.1046/j.1365-8711.2002.05899.x} {335, L89}

\bibitem[\protect\citeauthoryear{Johnston et~al.,}{Johnston
  et~al.}{2019}]{Johnston2019}
Johnston H.,  et~al., 2019, Astronomy \& Astrophysics, 624, A30

\bibitem[\protect\citeauthoryear{{Kiessling} et~al.,}{{Kiessling}
  et~al.}{2015}]{Kiessling2015}
{Kiessling} A.,  et~al., 2015, \mn@doi [\ssr] {10.1007/s11214-015-0203-6},
  \href {https://ui.adsabs.harvard.edu/abs/2015SSRv..193...67K} {193, 67}

\bibitem[\protect\citeauthoryear{{Kirk} et~al.,}{{Kirk}
  et~al.}{2015}]{Kirk2015}
{Kirk} D.,  et~al., 2015, \mn@doi [\ssr] {10.1007/s11214-015-0213-4}, \href
  {https://ui.adsabs.harvard.edu/abs/2015SSRv..193..139K} {193, 139}

\bibitem[\protect\citeauthoryear{{Lacerna} et~al.,}{{Lacerna}
  et~al.}{2022}]{Lacerna2022}
{Lacerna} I.,  et~al., 2022, \mn@doi [\mnras] {10.1093/mnras/stac1020}, \href
  {https://ui.adsabs.harvard.edu/abs/2022MNRAS.513.2271L} {513, 2271}

\bibitem[\protect\citeauthoryear{{Libeskind}, {Hoffman}, {Tully}, {Courtois},
  {Pomar{\`e}de}, {Gottl{\"o}ber}  \& {Steinmetz}}{{Libeskind}
  et~al.}{2015}]{Libeskind2015}
{Libeskind} N.~I.,  {Hoffman} Y.,  {Tully} R.~B.,  {Courtois} H.~M.,
  {Pomar{\`e}de} D.,  {Gottl{\"o}ber} S.,   {Steinmetz} M.,  2015, \mn@doi
  [\mnras] {10.1093/mnras/stv1302}, \href
  {https://ui.adsabs.harvard.edu/abs/2015MNRAS.452.1052L} {452, 1052}

\bibitem[\protect\citeauthoryear{{Maier}, {Haines}  \& {Ziegler}}{{Maier}
  et~al.}{2022}]{Maier2022}
{Maier} C.,  {Haines} C.~P.,   {Ziegler} B.~L.,  2022, \mn@doi [\aap]
  {10.1051/0004-6361/202141498}, \href
  {https://ui.adsabs.harvard.edu/abs/2022A&A...658A.190M} {658, A190}

\bibitem[\protect\citeauthoryear{{Nelson} et~al.,}{{Nelson}
  et~al.}{2019}]{Nelson2019}
{Nelson} D.,  et~al., 2019, \mn@doi [Computational Astrophysics and Cosmology]
  {10.1186/s40668-019-0028-x}, \href
  {https://ui.adsabs.harvard.edu/abs/2019ComAC...6....2N} {6, 2}

\bibitem[\protect\citeauthoryear{{Otter}, {Masters}, {Simmons}  \&
  {Lintott}}{{Otter} et~al.}{2020}]{Otter2020}
{Otter} J.~A.,  {Masters} K.~L.,  {Simmons} B.,   {Lintott} C.~J.,  2020,
  \mn@doi [\mnras] {10.1093/mnras/stz3626}, \href
  {https://ui.adsabs.harvard.edu/abs/2020MNRAS.492.2722O} {492, 2722}

\bibitem[\protect\citeauthoryear{{Pawlowski}}{{Pawlowski}}{2018}]{Pawlowski2018}
{Pawlowski} M.~S.,  2018, \mn@doi [Modern Physics Letters A]
  {10.1142/S0217732318300045}, \href
  {https://ui.adsabs.harvard.edu/abs/2018MPLA...3330004P} {33, 1830004}

\bibitem[\protect\citeauthoryear{Paz, Stasyszyn  \& Padilla}{Paz
  et~al.}{2008}]{paz2008angular}
Paz D.~J.,  Stasyszyn F.,   Padilla N.~D.,  2008, Monthly Notices of the Royal
  Astronomical Society, \href
  {https://doi.org/10.1111/j.1365-2966.2008.13655.x} {389, 1127}

\bibitem[\protect\citeauthoryear{Paz, Sgr{\'o}, Merch{\'a}n  \& Padilla}{Paz
  et~al.}{2011}]{paz2011alignments}
Paz D.~J.,  Sgr{\'o} M.~A.,  Merch{\'a}n M.,   Padilla N.,  2011, Monthly
  Notices of the Royal Astronomical Society, \href
  {https://doi.org/10.1111/j.1365-2966.2011.18518.x} {414, 2029}

\bibitem[\protect\citeauthoryear{{Peebles}}{{Peebles}}{1980}]{Peebles1980}
{Peebles} P.~J.~E.,  1980, {The large-scale structure of the universe}.
Princeton university press

\bibitem[\protect\citeauthoryear{Pen, Lee  \& Seljak}{Pen
  et~al.}{2000}]{Pen2000}
Pen U.-L.,  Lee J.,   Seljak U.,  2000, The Astrophysical Journal, \href
  {https://doi.org/10.1086/317273} {543, L107}

\bibitem[\protect\citeauthoryear{{Peschken} \& {{\L}okas}}{{Peschken} \&
  {{\L}okas}}{2019}]{Peschken2019}
{Peschken} N.,  {{\L}okas} E.~L.,  2019, \mn@doi [\mnras]
  {10.1093/mnras/sty3277}, \href
  {https://ui.adsabs.harvard.edu/abs/2019MNRAS.483.2721P} {483, 2721}

\bibitem[\protect\citeauthoryear{Planck~Collaboration Ade
  et~al.,}{Planck~Collaboration et~al.}{2016}]{Planck2016}
Planck~Collaboration Ade P.~A.,  et~al., 2016, Astronomy \& Astrophysics, \href
  {https://doi.org/10.1051/0004-6361/201525830} {594, A13}

\bibitem[\protect\citeauthoryear{Porciani, Dekel  \& Hoffman}{Porciani
  et~al.}{2002a}]{Porciani2002a}
Porciani C.,  Dekel A.,   Hoffman Y.,  2002a, Monthly Notices of the Royal
  Astronomical Society, \href
  {https://doi.org/10.1046/j.1365-8711.2002.05305.x} {332, 325}

\bibitem[\protect\citeauthoryear{Porciani, Dekel  \& Hoffman}{Porciani
  et~al.}{2002b}]{Porciani2002b}
Porciani C.,  Dekel A.,   Hoffman Y.,  2002b, Monthly Notices of the Royal
  Astronomical Society, \href
  {https://doi.org/10.1046/j.1365-8711.2002.05899.x} {332, 339}

\bibitem[\protect\citeauthoryear{{Ragone-Figueroa}, {Granato}, {Borgani}, {De
  Propris}, {Garc{\'\i}a Lambas}, {Murante}, {Rasia}  \&
  {West}}{{Ragone-Figueroa} et~al.}{2020}]{RagoneFigueroa2020}
{Ragone-Figueroa} C.,  {Granato} G.~L.,  {Borgani} S.,  {De Propris} R.,
  {Garc{\'\i}a Lambas} D.,  {Murante} G.,  {Rasia} E.,   {West} M.,  2020,
  \mn@doi [\mnras] {10.1093/mnras/staa1389}, \href
  {https://ui.adsabs.harvard.edu/abs/2020MNRAS.495.2436R} {495, 2436}

\bibitem[\protect\citeauthoryear{Rodriguez \& Merch{\'a}n}{Rodriguez \&
  Merch{\'a}n}{2020}]{rodriguez20}
Rodriguez F.,  Merch{\'a}n M.,  2020, Astronomy \& Astrophysics, \href
  {https://www.aanda.org/articles/aa/abs/2020/04/aa37423-19/aa37423-19.html}
  {636, A61}

\bibitem[\protect\citeauthoryear{{Rodriguez-Gomez} et~al.,}{{Rodriguez-Gomez}
  et~al.}{2015}]{Rodriguez-Gomez2015}
{Rodriguez-Gomez} V.,  et~al., 2015, \mn@doi [\mnras] {10.1093/mnras/stv264},
  \href {https://ui.adsabs.harvard.edu/abs/2015MNRAS.449...49R} {449, 49}

\bibitem[\protect\citeauthoryear{Rodriguez, Merch{\'a}n  \& Artale}{Rodriguez
  et~al.}{2022}]{Rodriguez2022}
Rodriguez F.,  Merch{\'a}n M.,   Artale M.~C.,  2022, Monthly Notices of the
  Royal Astronomical Society, 514, 1077

\bibitem[\protect\citeauthoryear{Sales \& Lambas}{Sales \&
  Lambas}{2004}]{Sales2004}
Sales L.,  Lambas D.~G.,  2004, Monthly Notices of the Royal Astronomical
  Society, \href {https://doi.org/10.1111/j.1365-2966.2004.07443.x} {348, 1236}

\bibitem[\protect\citeauthoryear{Samuroff, Mandelbaum  \& Di~Matteo}{Samuroff
  et~al.}{2020}]{Samuroff2020}
Samuroff S.,  Mandelbaum R.,   Di~Matteo T.,  2020, Monthly Notices of the
  Royal Astronomical Society, 491, 5330

\bibitem[\protect\citeauthoryear{{Shao}, {Cautun}, {Frenk}, {Gao}, {Crain},
  {Schaller}, {Schaye}  \& {Theuns}}{{Shao} et~al.}{2016}]{Shao2016}
{Shao} S.,  {Cautun} M.,  {Frenk} C.~S.,  {Gao} L.,  {Crain} R.~A.,  {Schaller}
  M.,  {Schaye} J.,   {Theuns} T.,  2016, \mn@doi [\mnras]
  {10.1093/mnras/stw1247}, \href
  {https://ui.adsabs.harvard.edu/abs/2016MNRAS.460.3772S} {460, 3772}

\bibitem[\protect\citeauthoryear{{Shi}, {Osato}, {Kurita}  \& {Takada}}{{Shi}
  et~al.}{2021a}]{Shi2021b}
{Shi} J.,  {Osato} K.,  {Kurita} T.,   {Takada} M.,  2021a, \mn@doi [\apj]
  {10.3847/1538-4357/ac0cfa}, \href
  {https://ui.adsabs.harvard.edu/abs/2021ApJ...917..109S} {917, 109}

\bibitem[\protect\citeauthoryear{{Shi}, {Kurita}, {Takada}, {Osato},
  {Kobayashi}  \& {Nishimichi}}{{Shi} et~al.}{2021b}]{Shi2021}
{Shi} J.,  {Kurita} T.,  {Takada} M.,  {Osato} K.,  {Kobayashi} Y.,
  {Nishimichi} T.,  2021b, \mn@doi [\jcap] {10.1088/1475-7516/2021/03/030},
  \href {https://ui.adsabs.harvard.edu/abs/2021JCAP...03..030S} {2021, 030}

\bibitem[\protect\citeauthoryear{{Springel}}{{Springel}}{2010}]{Springel2010}
{Springel} V.,  2010, \mn@doi [\mnras] {10.1111/j.1365-2966.2009.15715.x},
  \href {https://ui.adsabs.harvard.edu/abs/2010MNRAS.401..791S} {401, 791}

\bibitem[\protect\citeauthoryear{{Springel}, {White}, {Tormen}  \&
  {Kauffmann}}{{Springel} et~al.}{2001}]{Springel2001}
{Springel} V.,  {White} S. D.~M.,  {Tormen} G.,   {Kauffmann} G.,  2001,
  \mn@doi [\mnras] {10.1046/j.1365-8711.2001.04912.x}, \href
  {https://ui.adsabs.harvard.edu/abs/2001MNRAS.328..726S} {328, 726}

\bibitem[\protect\citeauthoryear{{Stott}}{{Stott}}{2022}]{Stott2022}
{Stott} J.~P.,  2022, \mn@doi [\mnras] {10.1093/mnras/stac089}, \href
  {https://ui.adsabs.harvard.edu/abs/2022MNRAS.511.2659S} {511, 2659}

\bibitem[\protect\citeauthoryear{{Tenneti}, {Kitching}, {Joachimi}  \& {Di
  Matteo}}{{Tenneti} et~al.}{2021}]{Tenneti2021}
{Tenneti} A.,  {Kitching} T.~D.,  {Joachimi} B.,   {Di Matteo} T.,  2021,
  \mn@doi [\mnras] {10.1093/mnras/staa3934}, \href
  {https://ui.adsabs.harvard.edu/abs/2021MNRAS.501.5859T} {501, 5859}

\bibitem[\protect\citeauthoryear{Usami \& Fujimoto}{Usami \&
  Fujimoto}{1997}]{Usami1997}
Usami M.,  Fujimoto M.,  1997, The Astrophysical Journal, \href
  {https://doi.org/10.1086/304624} {487, 489}

\bibitem[\protect\citeauthoryear{{Vogelsberger} et~al.,}{{Vogelsberger}
  et~al.}{2014a}]{Vogelsberger2014a}
{Vogelsberger} M.,  et~al., 2014a, \mn@doi [\mnras] {10.1093/mnras/stu1536},
  \href {https://ui.adsabs.harvard.edu/abs/2014MNRAS.444.1518V} {444, 1518}

\bibitem[\protect\citeauthoryear{{Vogelsberger} et~al.,}{{Vogelsberger}
  et~al.}{2014b}]{Vogelsberger2014b}
{Vogelsberger} M.,  et~al., 2014b, \mn@doi [\nat] {10.1038/nature13316}, \href
  {https://ui.adsabs.harvard.edu/abs/2014Natur.509..177V} {509, 177}

\bibitem[\protect\citeauthoryear{{Wang}, {Yang}, {Mo}, {Li}, {van den Bosch},
  {Fan}  \& {Chen}}{{Wang} et~al.}{2008}]{Wang2008}
{Wang} Y.,  {Yang} X.,  {Mo} H.~J.,  {Li} C.,  {van den Bosch} F.~C.,  {Fan}
  Z.,   {Chen} X.,  2008, \mn@doi [\mnras] {10.1111/j.1365-2966.2008.12927.x},
  \href {https://ui.adsabs.harvard.edu/abs/2008MNRAS.385.1511W} {385, 1511}

\bibitem[\protect\citeauthoryear{{Welker}, {Dubois}, {Pichon}, {Devriendt}  \&
  {Chisari}}{{Welker} et~al.}{2018}]{Welker2018}
{Welker} C.,  {Dubois} Y.,  {Pichon} C.,  {Devriendt} J.,   {Chisari} N.~E.,
  2018, \mn@doi [\aap] {10.1051/0004-6361/201629007}, \href
  {https://ui.adsabs.harvard.edu/abs/2018A&A...613A...4W} {613, A4}

\bibitem[\protect\citeauthoryear{Yang, Mo, Van Den~Bosch  \& Jing}{Yang
  et~al.}{2005}]{Yang2005}
Yang X.,  Mo H.,  Van Den~Bosch F.~C.,   Jing Y.,  2005, Monthly Notices of the
  Royal Astronomical Society, 356, 1293

\bibitem[\protect\citeauthoryear{Yang, Van Den~Bosch, Mo, Mao, Kang, Weinmann,
  Guo  \& Jing}{Yang et~al.}{2006}]{yang2006alignment}
Yang X.,  Van Den~Bosch F.~C.,  Mo H.,  Mao S.,  Kang X.,  Weinmann S.~M.,  Guo
  Y.,   Jing Y.,  2006, Monthly Notices of the Royal Astronomical Society,
  \href {https://doi.org/10.1111/j.1365-2966.2006.10373.x} {369, 1293}

\bibitem[\protect\citeauthoryear{{Zhang}, {Lee}, {Krolewski}, {Shi}, {Horowitz}
   \& {Kooistra}}{{Zhang} et~al.}{2022}]{Zhang2022}
{Zhang} B.,  {Lee} K.-G.,  {Krolewski} A.,  {Shi} J.,  {Horowitz} B.,
  {Kooistra} R.,  2022, arXiv e-prints, \href
  {https://ui.adsabs.harvard.edu/abs/2022arXiv221109331Z} {p. arXiv:2211.09331}

\bibitem[\protect\citeauthoryear{{Zjupa}, {Sch{\"a}fer}  \& {Hahn}}{{Zjupa}
  et~al.}{2020}]{Zjupa2020}
{Zjupa} J.,  {Sch{\"a}fer} B.~M.,   {Hahn} O.,  2020, arXiv e-prints, \href
  {https://ui.adsabs.harvard.edu/abs/2020arXiv201007951Z} {p. arXiv:2010.07951}

\makeatother
\end{thebibliography}



\bsp	
\label{lastpage}
\end{document}